\documentclass[12pt]{article}
\usepackage{amssymb}
\usepackage{amsfonts,amssymb,eucal,amsmath,epsfig}
\title{Planar Symmetric Concave Central Configurations in Four-body Problem}
\author{{\small\sc Chunhua Deng$^1$ and Shiqing Zhang$^2$}\\
\small 1. Faculty of Mathematics and Physics, Huaiyin Institute of Technology, Huai'an 223003, China\\
\small chdeng8011@sohu.com\\
\small 2. College of Mathematics, Sichuan University, Chengdu
610064, China}
\date{}

\begin{document}
\maketitle

Abstract: In this paper, we consider the problem: given a symmetric
concave configuration of four bodies, under what conditions is it
possible to choose positive masses which make it central. We show
that there are some regions in which no central configuration is
possible for positive masses. Conversely, for any configuration in
the complement of the union of these regions, it is always possible
to choose positive masses
to make the configuration central.\\


\section*{1. Introduction and Main Results}
\setcounter{section}{1} \setcounter{equation}{0}

The Newtonian $n$-body problem concerns the motion of $n$ mass
points with masses $m_i\in \mathbb{R}^{+}, i=1,2,\cdots,n$. The
motion is governed by Newton's law of gravitation:
\begin{equation}
m_i\ddot{q}_i=\sum\limits_{k\neq
i}\frac{m_km_i(q_k-q_i)}{|q_k-q_i|^3}, i=1,2,\cdots,n,
\end{equation}
where $q_i\in \mathbb{R}^d (d=1,2,3)$ is the  position of $m_i$.
Alternatively the system (1.1) can be written
\begin{equation}
m_i\ddot{q}_i=\frac{\partial U(q)}{\partial q_i}, i=1,2,\cdots,n
\end{equation}
where
\begin{equation}
U(q)=U(q_1,q_2,\cdots,q_n)=\sum\limits_{1\leq k<j\leq
n}\frac{m_km_j}{|q_k-q_j|}
\end{equation}
is the Newtonian potential of system (1.1). Let
$$C=m_1q_1+\cdots+m_nq_n, M=m_1+\cdots+m_n, c=C/M$$
be the first moment, total mass and center of mass of the bodies,
respectively. The set $\bigtriangleup$ of collision configurations
is defined by
$$\bigtriangleup=\{q\in (\mathbb{R}^d)^n: q_i=q_j\;\text{for some}\; i \neq j \}.$$
A configuration $q =(q_1,\cdots,q_n)\in (\mathbb{R}^d)^n
\backslash\bigtriangleup$ is called a central configuration if there
exists some positive constant $\lambda$ such that

\begin{equation}
-\lambda (q_i-c)=\sum\limits_{j=1,j\neq
i}^{n}\frac{m_j(q_j-q_i)}{|q_j-q_i|^3},\quad i=1,2,\cdots,n.
\end{equation}
Furthermore it can be easily verified that $\lambda=U/I$, where $I$
is the moment of inertial of the system, i.e.
$I=\sum\limits_{i=1}^{N}m_i|q_i|^2$. The set of central
configurations are invariant under three classes of transformations
on $(\mathbb{R}^d)^n$: translations, scalings, and orthogonal
transformations.
 A configuration $q = (q_1,\cdots,q_n)$
is concave if one mass point is in the interior of the triangle
formed by the other three mass points. For $n=4,q_i\in
\mathbb{R}^2$, Long and Sun [] proved

\noindent{\bf Lemma 1.1.} Let $\alpha,\beta>0$ be any two given real
numbers. Let $q = (q_1,q_2,q_3,q_4)\in (\mathbb{R}^2)^4$ be a
concave non-collinear central configuration with masses
$(\beta,\alpha,\beta,\beta)$ respectively, and with $q_2$ located
inside the triangle formed by $q_1, q_3$, and $q_4$. Then the
configuration $q$ must possess a symmetry, so either $q_1, q_3$, and
$q_4$ form an equilateral triangle and $q_2$ is located at the
center of the triangle, or $q_1, q_3$, and $q_4$ form an isosceles
triangle, and $q_2$ is on the symmetrical axis of the triangle.

In this paper we consider the inverse problem: given a planar
symmetric concave configuration (Figure 1), find the positive mass
vectors, if any, for which it is a central configuration. The
equations for the central configurations can be written as
\begin{equation}
\left\{
\begin{array}{l}
m_2\frac{q_2-q_1}{|q_2-q_1|^3}+m_3\frac{q_3-q_1}{|q_3-q_1|^3}+m_4\frac{q_4-q_1}{|q_4-q_1|^3}=-\lambda(q_1-c)\\
m_1\frac{q_1-q_2}{|q_1-q_2|^3}+m_3\frac{q_3-q_2}{|q_3-q_2|^3}+m_4\frac{q_4-q_2}{|q_4-q_2|^3}=-\lambda(q_2-c)\\
m_1\frac{q_1-q_3}{|q_1-q_3|^3}+m_2\frac{q_2-q_3}{|q_2-q_3|^3}+m_4\frac{q_4-q_3}{|q_4-q_3|^3}=-\lambda(q_3-c)\\
m_1\frac{q_1-q_4}{|q_1-q_4|^3}+m_2\frac{q_2-q_4}{|q_2-q_4|^3}+m_3\frac{q_3-q_4}{|q_3-q_4|^3}=-\lambda(q_4-c)\\
\end{array}
\right.
\end{equation}

We can obtain the following results:

\noindent{\bf Theorem 1.1.} Let $q_1=(-1,0)$, $q_2=(1,0)$,
$q_3=(0,t)$, $q_4=(0,s)$ where $t>s>0$, and assume that the center
of mass $c=C/M=q_4$. The symmetric concave configuration $q=(q_1,
q_2, q_3, q_4)$ can be a central configuration if and only if
$t=\sqrt{3}, s=\frac{\sqrt{3}}{3}$, and  the masses of $q_1$, $q_2$
and $q_3$ are all equal, i.e. $m_1=m_2=m_3>0$. The mass of $q_4$ can
be any positive number $m_4>0$.

\noindent{\bf Theorem 1.2.} Let $q_1=(-1,0)$, $q_2=(1,0)$,
$q_3=(0,t)$, $q_4=(0,s)$, where $t>s>0$, and assume that the center
of mass $c=C/M\neq q_4$. There exists two open bounded regions $C$
and $D$ which can be seen in figure (), the configuration
$q=(q_1,q_2,q_3,q_4)$ can be a central configuration with positive
masses, where
\begin{equation}
\begin{aligned}
 m_1=m_2=\lambda\frac{2^3\sqrt{1+t^2}^3(t-c_y)}{2t\sqrt{1+s^2}^3(t-s)^3}\frac{(t-s)^3-\sqrt{1+s^2}^3}{\frac{2}{\sqrt{1+s^2}})^3-(\frac{\sqrt{1+t^2}}{t-s})^3}\\
 \end{aligned}
\end{equation}

\begin{equation}
\begin{aligned}
 m_3=\frac{\lambda s\sqrt{1+t^2}^3}{\sqrt{1+s^2}^6(t-s)^3}\frac{(\sqrt{1+s^2}^3-2^3)(\sqrt{1+s^2}^3-(t-s)^3)}{(\frac{t-s}{(t-s)^3}+\frac{s}{\sqrt{1+s^2}^3}-\frac{t}{\sqrt{1+t^2}^3})((\frac{2}{\sqrt{1+s^2}})^3-(\frac{\sqrt{1+t^2}}{t-s})^3)}\\
 \end{aligned}
\end{equation}
\begin{equation}
m_4=\frac{\lambda(t-c_y)}{(t-s)}\frac{(2^3-\sqrt{1+t^2}^3)}{((\frac{2}{\sqrt{1+s^2}})^3-(\frac{\sqrt{1+t^2}}{t-s})^3)}.
\end{equation}
\begin{figure}[htb]
\begin{center}
\includegraphics[width=10cm]{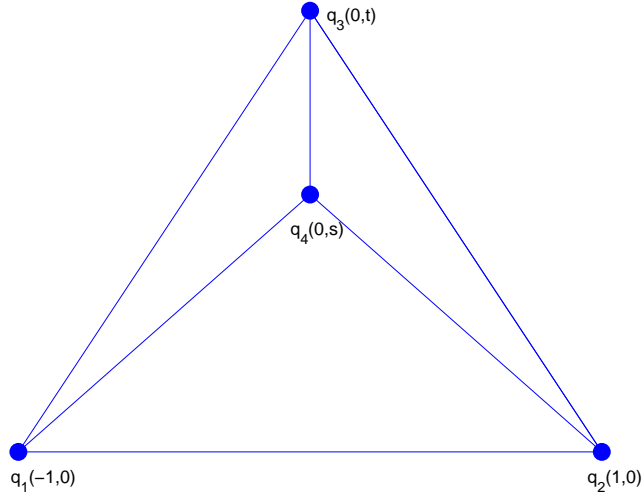}
\caption{\label{TheLabel}The symmetric concave configuration}
\end{center}
\end{figure}
\section*{2. General Symmetric Concave Central Configurations with Four bodies}
\setcounter{section}{2} \setcounter{equation}{0}

Assume the center of mass $c=(c_x,c_y)$. Given $q_1=(-1,0)$,
$q_2=(1,0)$, $q_3=(0,t)$, $q_4=(0,s)$, where $t>s>0$, the system
(1.5) can be divided into two parts:
\begin{equation}
\left\{
\begin{array}{l}
\frac{2}{2^3}m_2+\frac{1}{\sqrt{1+t^2}^3}m_3+\frac{1}{\sqrt{1+s^2}^3}m_4=\lambda(1+c_x)\\
\frac{-2}{2^3}m_2+\frac{-1}{\sqrt{1+t^2}^3}m_3+\frac{-1}{\sqrt{1+s^2}^3}m_4=-\lambda(1-c_x)\\
\frac{-1}{\sqrt{1+t^2}^3}m_1+\frac{1}{\sqrt{1+t^2}^3}m_2=\lambda c_x\\
\frac{-1}{\sqrt{1+s^2}^3}m_1+\frac{1}{\sqrt{1+s^2}^3}m_2=\lambda c_x\\
\end{array}
\right.
\end{equation}
and
\begin{equation}
\left\{
\begin{array}{l}
\frac{t}{\sqrt{1+t^2}^3}m_3+\frac{s}{\sqrt{1+s^2}^3}m_4=\lambda c_y\\
\frac{t}{\sqrt{1+t^2}^3}m_3+\frac{s}{\sqrt{1+s^2}^3}m_4=\lambda c_y\\
\frac{-t}{\sqrt{1+t^2}^3}m_1+\frac{-t}{\sqrt{1+t^2}^3}m_2+\frac{s-t}{(t-s)^3}m_4=-\lambda(t-c_y)\\
\frac{-s}{\sqrt{1+s^2}^3}m_1+\frac{-s}{\sqrt{1+s^2}^3}m_2+\frac{t-s}{(t-s)^3}m_3=-\lambda(s-c_y).\\
\end{array}
\right.
\end{equation}
In (2.2) the first two equations are identical. The third and the
fourth equations in (2.1) imply that
$$(\frac{1}{\sqrt{1+t^2}^3}-\frac{1}{\sqrt{1+s^2}^3})(m_2-m_1)=0.$$
For $t>s>0$ we have
$$m_1=m_2.$$
The first two equations in (2.1) together with $m_1=m_2$ and
positive number $\lambda>0$ imply that
$$c_x=0.$$
Thus systems (1.5) for central configurations become
\begin{equation}
\left\{
\begin{array}{l}
\frac{2}{2^3}m_2+\frac{1}{\sqrt{1+t^2}^3}m_3+\frac{1}{\sqrt{1+s^2}^3}m_4=\lambda\\
\frac{t}{\sqrt{1+t^2}^3}m_3+\frac{s}{\sqrt{1+s^2}^3}m_4=\lambda c_y\\
\frac{-2t}{\sqrt{1+t^2}^3}m_2+\frac{s-t}{(t-s)^3}m_4=-\lambda(t-c_y)\\
\frac{-2s}{\sqrt{1+s^2}^3}m_2+\frac{t-s}{(t-s)^3}m_3=-\lambda(s-c_y).\\
\end{array}
\right.
\end{equation}

\section*{3. The Proof of Theorem 1.1}
\setcounter{section}{3} \setcounter{equation}{0}

In this section, we will find the solution of masses $m_1, m_2, m_3,
m_4$ with two parameters $s, t$ for the four-body central
configuration. We assume the center of mass $c=C/M=q_4$, i.e.
$c_y=s$. the system (2.3) for central configurations become
\begin{equation}
\left\{
\begin{array}{l}
\frac{2}{2^3}m_2+\frac{1}{\sqrt{1+t^2}^3}m_3+\frac{1}{\sqrt{1+s^2}^3}m_4=\lambda\\
\frac{t}{\sqrt{1+t^2}^3}m_3+\frac{s}{\sqrt{1+s^2}^3}m_4=\lambda s\\
\frac{-2t}{\sqrt{1+t^2}^3}m_2+\frac{s-t}{(t-s)^3}m_4=-\lambda(t-s)\\
\frac{-2s}{\sqrt{1+s^2}^3}m_2+\frac{t-s}{(t-s)^3}m_3=0.\\
\end{array}
\right.
\end{equation}
The fourth equation in (3.1) can be written
\begin{equation}
m_2=\frac{t-s}{(t-s)^3}\frac{\sqrt{1+s^2}^3}{2s}m_3.
\end{equation}
Substituting (3.2) into the third equation in (3.1), we have
\begin{equation}
\frac{-2t}{\sqrt{1+t^2}^3}\frac{t-s}{(t-s)^3}\frac{\sqrt{1+s^2}^3}{2s}m_3+\frac{s-t}{(t-s)^3}m_4=-\lambda
(t-s),
\end{equation}
for $t>s>0$, then
\begin{equation}
\frac{t}{\sqrt{1+t^2}^3}\frac{1}{(t-s)^3}\frac{\sqrt{1+s^2}^3}{s}m_3+\frac{1}{(t-s)^3}m_4=\lambda.
\end{equation}
From he second equation in (3.1) and the above equation (3.4), we
have
\begin{equation}
(t-s)=\sqrt{1+s^2}.
\end{equation}
Thus the last three equations in (3.1) is equivalent to
\begin{equation}
\left\{
\begin{array}{l}
(t-s)=\sqrt{1+s^2}\\
m_4=\lambda \sqrt{1+s^2}^3-\frac{2t}{\sqrt{1+t^2}^3}\frac{\sqrt{1+s^2}^3}{t-s}m_2\\
m_3=\frac{2s}{t-s}m_2.\\
\end{array}
\right.
\end{equation}
Substituting (3.6) into the first equation in (3.1) and simplifying,
we have
\begin{equation}
t=\sqrt{3}, s=\frac{\sqrt{3}}{3}.
\end{equation}
Then we have $m_1=m_2=m_3$ and
$m_4=\frac{8}{9}\sqrt{3}\lambda-\frac{\sqrt{3}}{3}m_2$. Furthermore,
for any positive mass $m_4>0$, we can choose suitable $\lambda>0$
such that $m_4=\frac{8}{9}\sqrt{3}\lambda-\frac{\sqrt{3}}{3}m_2$.
This completes the proof of Theorem 1.1.
\section*{4. The Proof of Theorem 1.2}
\setcounter{section}{4} \setcounter{equation}{0}

In this section, we assume the center of mass $c=C/M\neq q_4$, i.e.
$c_y\neq s$. Combining the second and the third equations in (2.3)
and eliminating $m_4$, the following equation is derived
\begin{equation}
\frac{2t}{\sqrt{1+t^2}^3}\frac{s}{\sqrt{1+s^2}^3}m_2+\frac{t}{\sqrt{1+t^2}^3}\frac{s-t}{(t-s)^3}m_3=\lambda((\frac{s-t}{(t-s)^3}-\frac{s}{\sqrt{1+s^2}^3})c_y+\frac{ts}{\sqrt{1+s^2}^3}).
\end{equation}
Multiplying both sides of the fourth equation in (2.3) by
$\frac{t}{\sqrt{1+t^2}^3}$, we have
\begin{equation}
\frac{2t}{\sqrt{1+t^2}^3}\frac{s}{\sqrt{1+s^2}^3}m_2+\frac{t}{\sqrt{1+t^2}^3}\frac{s-t}{(t-s)^3}m_3=\lambda(c_y-s)\frac{t}{\sqrt{1+t^2}^3}.
\end{equation}
Then we have the necessary conditions for the solvability of (2.3):
\begin{equation}
(\frac{s-t}{(t-s)^3}-\frac{s}{\sqrt{1+s^2}^3})c_y+\frac{ts}{\sqrt{1+s^2}^3}=(c_y-s)\frac{t}{\sqrt{1+t^2}^3},
\end{equation}
then
\begin{equation}
c_y=(\frac{ts}{\sqrt{1+s^2}^3}-\frac{ts}{\sqrt{1+t^2}^3})
/(\frac{t-s}{(t-s)^3}+\frac{s}{\sqrt{1+s^2}^3}-\frac{t}{\sqrt{1+t^2}^3}).
\end{equation}
The system (2.3) for central configurations  become
\begin{equation}
\left\{
\begin{array}{l}
c_y=(\frac{ts}{\sqrt{1+s^2}^3}-\frac{ts}{\sqrt{1+t^2}^3})
/(\frac{t-s}{(t-s)^3}+\frac{s}{\sqrt{1+s^2}^3}-\frac{t}{\sqrt{1+t^2}^3})\\
\frac{2}{2^3}m_2+\frac{1}{\sqrt{1+t^2}^3}m_3+\frac{1}{\sqrt{1+s^2}^3}m_4=\lambda\\
\frac{t}{\sqrt{1+t^2}^3}m_3+\frac{s}{\sqrt{1+s^2}^3}m_4=\lambda c_y\\
\frac{-2t}{\sqrt{1+t^2}^3}m_2+\frac{s-t}{(t-s)^3}m_4=-\lambda(t-c_y).\\
\end{array}
\right.
\end{equation}

The third and the fourth equations in (4.5) can be written
\begin{equation}
\frac{1}{\sqrt{1+t^2}^3}m_3=\frac{1}{t}(\lambda
c_y-\frac{s}{\sqrt{1+s^2}^3}m_4),
\end{equation}
\begin{equation}
\frac{2}{2^3}m_2=(\lambda(t-c_y)-\frac{t-s}{(t-s)^3}m_4)\frac{\sqrt{1+t^2}^3}{2t}\frac{2}{2^3},
\end{equation}
and substituting the above two equations into the second equation in
(4.5), we obtain
$$(\lambda(t-c_y)-\frac{t-s}{(t-s)^3}m_4)\frac{\sqrt{1+t^2}^3}{2t}\frac{2}{2^3}+\frac{1}{t}(\lambda
c_y-\frac{s}{\sqrt{1+s^2}^3}m_4)+\frac{1}{\sqrt{1+s^2}^3}m_4=\lambda,$$
then
\begin{equation}
m_4=\frac{\lambda(t-c_y)}{(t-s)}\frac{(2^3-\sqrt{1+t^2}^3)}{((\frac{2}{\sqrt{1+s^2}})^3-(\frac{\sqrt{1+t^2}}{t-s})^3)}.
\end{equation}
Substituting (4.8) into (4.6) and simplifying, we have
\begin{equation}
\begin{aligned}
 m_3=&\frac{\sqrt{1+t^2}^3}{t}(\lambda
c_y-\frac{s}{\sqrt{1+s^2}^3}m_4)\\
=&\frac{\lambda s\sqrt{1+t^2}^3}{\sqrt{1+s^2}^6(t-s)^3}\frac{(\sqrt{1+s^2}^3-2^3)(\sqrt{1+s^2}^3-(t-s)^3)}{(\frac{t-s}{(t-s)^3}+\frac{s}{\sqrt{1+s^2}^3}-\frac{t}{\sqrt{1+t^2}^3})((\frac{2}{\sqrt{1+s^2}})^3-(\frac{\sqrt{1+t^2}}{t-s})^3)}.\\
 \end{aligned}
\end{equation}
Substituting (4.8) into (4.7) and simplifying, we have
\begin{equation}
\begin{aligned}
 m_2=&(\lambda(t-c_y)-\frac{t-s}{(t-s)^3}m_4)\frac{\sqrt{1+t^2}^3}{2t}\\
=&\lambda\frac{2^3\sqrt{1+t^2}^3(t-c_y)}{2t\sqrt{1+s^2}^3(t-s)^3}\frac{((t-s)^3-\sqrt{1+s^2}^3)}{((\frac{2}{\sqrt{1+s^2}})^3-(\frac{\sqrt{1+t^2}}{t-s})^3)}.\\
 \end{aligned}
\end{equation}
\begin{equation}
 m_1=m_2=\lambda\frac{2^3\sqrt{1+t^2}^3(t-c_y)}{2t\sqrt{1+s^2}^3(t-s)^3}\frac{((t-s)^3-\sqrt{1+s^2}^3)}{((\frac{2}{\sqrt{1+s^2}})^3-(\frac{\sqrt{1+t^2}}{t-s})^3)}.
\end{equation}
Thus we give the necessary condition (4.4) for the existence of the
solution of masses, and give the solution of masses explicitly in
(4.8-4.11).
 In the following we will analyze the mass functions and find the
 possible region in $st$-plane such that the mass functions are
 positive.

\noindent{\bf Lemma 4.1.} The region in which $m_4>0$ for $t>s>0$ is
the union of $A$ and $B$ in figure 2 surrounded by curves
$t=\sqrt{3}$, $2(t-s)-\sqrt{1+t^2}\sqrt{1+s^2}=0$ and $t-s=0$.

{\bf Proof.} With simple computation, we can find the center of mass
\begin{equation}
c=(c_x,c_y)=(0,\frac{sm_4+tm_3}{m_1+m_2+m_3+m_4}),
\end{equation}
then $t-c_y>0$ for $t>s>0$. For convenience, we denote
$p_1=2^3-\sqrt{1+t^2}^3$ and
$p_2=(\frac{2}{\sqrt{1+s^2}})^3-(\frac{\sqrt{1+t^2}}{t-s})^3$. Thus
$m_4>0$ is equivalent to $\frac{p_1}{p_2}>0$. We can show that
$p_2=0$ give rise a smooth monotone increasing curve above the curve
$t=s$, and bounded from right by $s=\sqrt{3}$.
\begin{figure}[htb]
\begin{center}
\includegraphics[width=14cm]{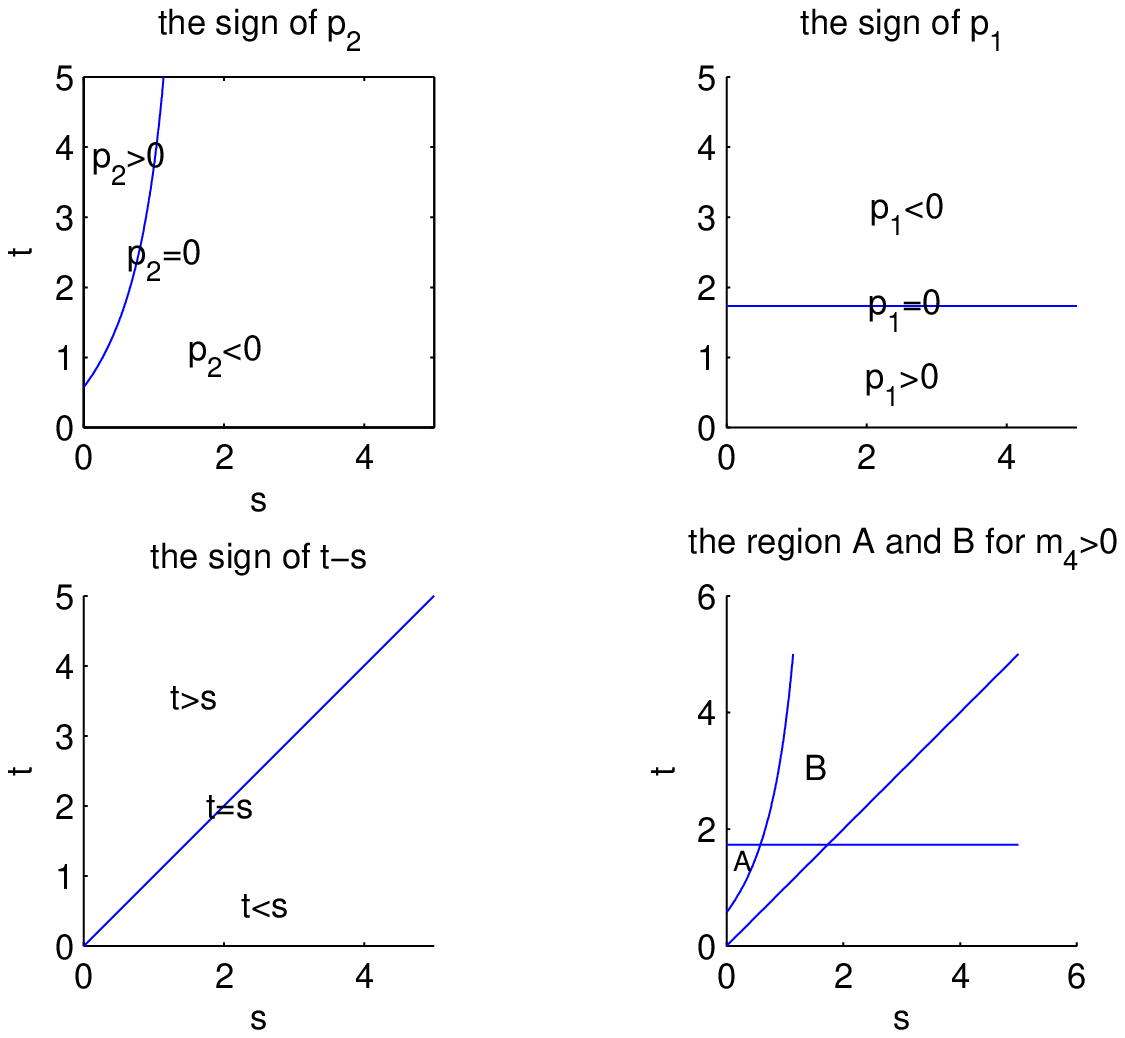}
\caption{\label{TheLabel}}
\end{center}
\end{figure}
$p_2=0$ is equivalent to $\sqrt{1+s^2}\sqrt{1+t^2}=2(t-s)$. We
observe that
$$\sqrt{1+s^2}\sqrt{1+t^2}=2(t-s)<2t,$$
then
$$\sqrt{1+s^2} < \frac{2t}{\sqrt{1+t^2}}< 2,$$
thus
 \begin{equation}
 s< \sqrt{3}.
\end{equation}
 Furthermore, from
$\sqrt{1+s^2}\sqrt{1+t^2}=2(t-s)$, we have
$$\lim_{t \to +\infty} 2(1-\frac{s}{t})=\lim_{t \to +\infty}\sqrt{1+s^2}\sqrt{1+\frac{1}{t^2}},$$
then
\begin{equation}
\lim_{t \to +\infty} s=\sqrt{3}.
\end{equation}
Let's take the derivative of $\sqrt{1+s^2}\sqrt{1+t^2}=2(t-s)$ with
respect to $s$,
$$(2-\frac{t\sqrt{1+s^2}}{\sqrt{1+t^2}}) \frac{dt}{ds}=2+\frac{s\sqrt{1+t^2}}{\sqrt{1+s^2}}.$$
Since
$$2-\frac{t\sqrt{1+s^2}}{\sqrt{1+t^2}}>2-\frac{2t}{\sqrt{1+t^2}}>0,$$
we have $$\frac{dt}{ds}>0.$$

Also the signs of $p_1, p_2$ are shown in the first three pictures
of Figure 2. So the region of $m_4>0$ is the union of two nonempty
open sets $A, B$ indicated in the fourth picture of Figure 2.
\begin{figure}[htb]
\begin{center}
\includegraphics[width=10cm]{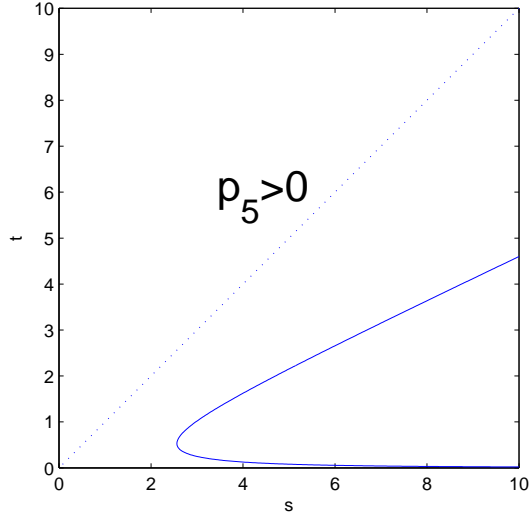}
\caption{\label{TheLabel}the sign of $p_5$}
\end{center}
\end{figure}

\noindent{\bf Lemma 4.2.} The region in which $m_4, m_3>0$ for
$t>s>0$ is the union of $C$ and $D$ in figure 3.

{\bf Proof.} For convenience, we denote $$p_3=\sqrt{1+s^2}^3-2^3,$$
$$p_4=\sqrt{1+s^2}^3-(t-s)^3,$$
$$p_5=\frac{t-s}{(t-s)^3}+\frac{s}{\sqrt{1+s^2}^3}-\frac{t}{\sqrt{1+t^2}^3}.$$
Then $m_3>0$ is equivalent to $\frac{p_3p_4}{p_5p_2}>0$.

By $t>s$ and $t-s<\sqrt{1+t^2}$, we have
\begin{equation}
\begin{aligned}
p_5= \frac{t-s}{(t-s)^3}+\frac{s}{\sqrt{1+s^2}^3}-\frac{t}{\sqrt{1+t^2}^3}=&t(\frac{1}{(t-s)^3}-\frac{1}{\sqrt{1+t^2}^3})+s(\frac{1}{\sqrt{1+s^2}^3}-\frac{1}{(t-s)^3})\\
>&s(\frac{1}{(t-s)^3}-\frac{1}{\sqrt{1+t^2}^3})+s(\frac{1}{\sqrt{1+s^2}^3}-\frac{1}{(t-s)^3})\\
=&s(\frac{1}{\sqrt{1+s^2}^3}-\frac{1}{\sqrt{1+t^2}^3})\\
>&0.\\
 \end{aligned}
\end{equation}

\begin{figure}[htb]
\begin{center}
\includegraphics[width=10cm]{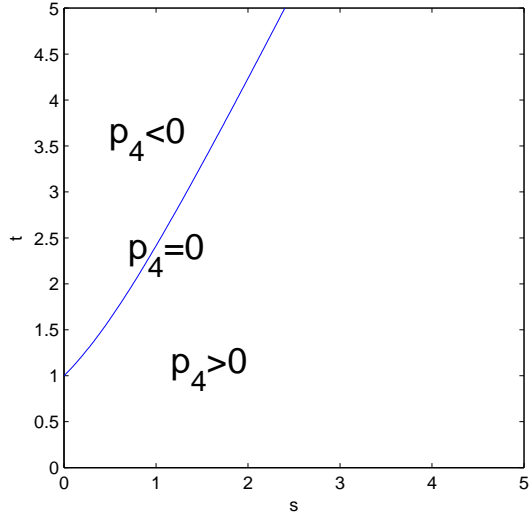}
\caption{\label{TheLabel}the sign of $p_4$}
\end{center}
\end{figure}
\begin{figure}[htb]
\begin{center}
\includegraphics[width=10cm]{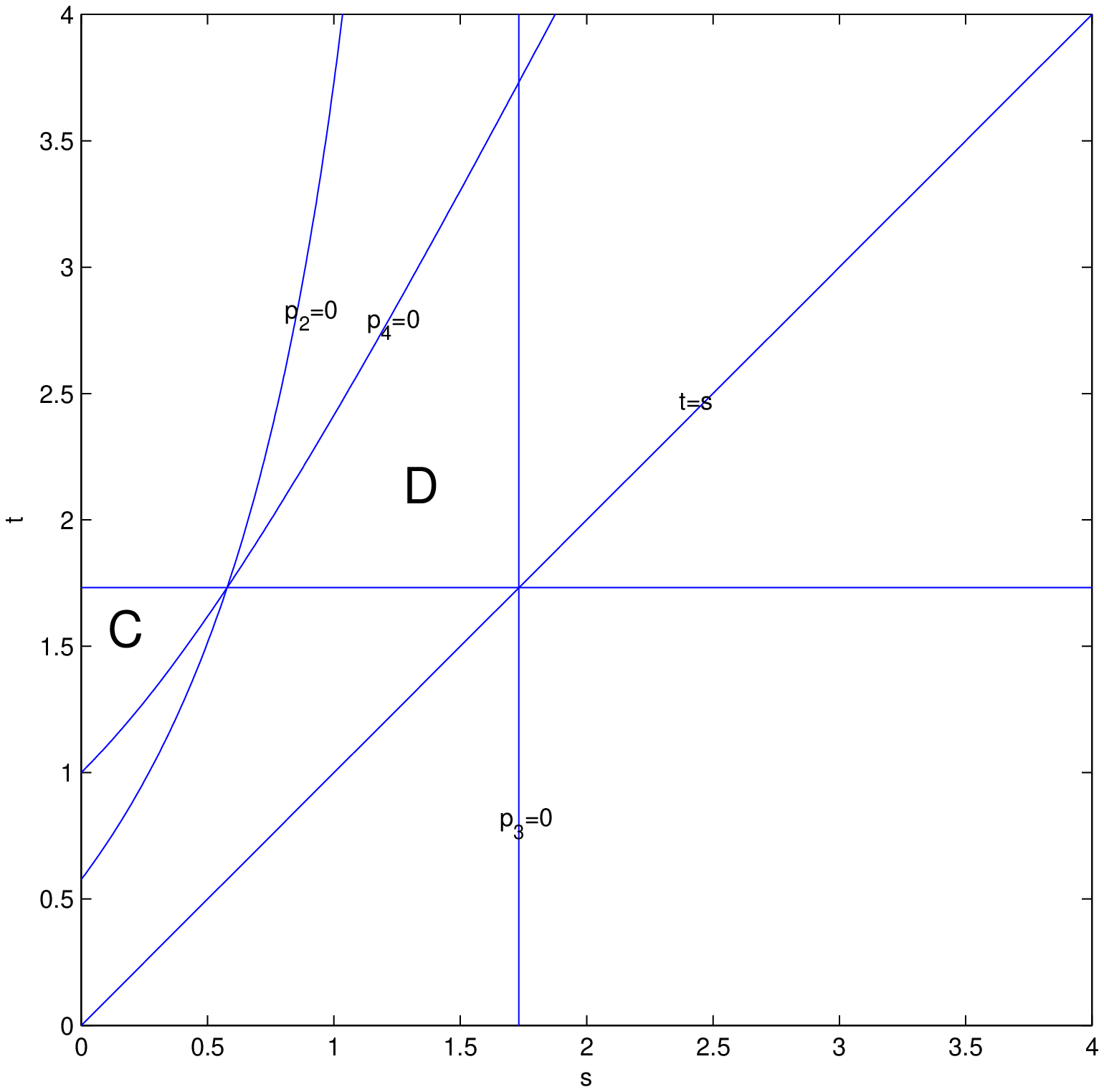}
\caption{\label{TheLabel}}
\end{center}
\end{figure}

The equation $p_3=0$ gives rise to a straight line $s=\sqrt{3}$ in
the $st$-plane. Also $p_3$ is positive on the right of this line.
The equation $p_4=0$ determines a smooth monotone increasing curve
$t=s+\sqrt{1+t^2}$, and $p_4$ is negative above this curve (Figure
4). With simple computation, we can find the implicit curves
$p_1=0$, $p_2=0$ and $p_4=0$ have only one intersecting point
$(\frac{\sqrt{3}}{3},3)$ with the domain $t>s>0$ which can be shown
in Figure 5. So the region of $m_4, m_3>0$ is the union of two
nonempty open sets $C, D$ indicated in Figure 5.

\noindent{\bf Lemma 4.3.} The region in which $m_i>0, i=1, 2, 3, 4$
for $t>s>0$ is just the union of $C$ and $D$ in figure 5.

{\bf Proof.} We have obtained
\begin{equation}
\begin{aligned}
 m_1=m_2=&\lambda\frac{2^3\sqrt{1+t^2}^3(t-c_y)}{2t\sqrt{1+s^2}^3(t-s)^3}\frac{((t-s)^3-\sqrt{1+s^2}^3)}{((\frac{2}{\sqrt{1+s^2}})^3-(\frac{\sqrt{1+t^2}}{t-s})^3)}\\
 =&-\lambda\frac{2^3\sqrt{1+t^2}^3(t-c_y)}{2t\sqrt{1+s^2}^3(t-s)^3}\frac{p_4}{p_2}\\
 \end{aligned}
 \end{equation}

The signs of $p_4, p_2$ decide the sign of $m_i, i=1,2$. This
complete the proof of Lemma 4.3.
\section*{Acknowledgements}
%The authors sincerely thank the referees and the editor for their many valuable
%comments which helped us improve the paper both in the content and
%also in the form.
Both authors are supported by NSFC, and the first
author is supported by the the Scientific Research Foundation of
Huaiyin Institute of Technology (HGA1102,HGB1004).

\end{document}